\documentclass[twocolumn,showpacs,aps,prd,superscriptaddress]{revtex4}
\pdfoutput = 1
\usepackage{graphicx}
\usepackage{dcolumn}
\usepackage{amsmath}

\begin{document}

\title{
  {\large 
    \bf \boldmath 
    Observing $CP$ Violation in Many-Body Decays
  }
}

\author{Mike Williams}
\affiliation{Physics Department, Imperial College London, London, SW7 2AZ, United Kingdom}

\date{\today}
\begin{abstract}
  It is well known that observing $CP$ violation in many-body decays could provide strong evidence for physics beyond the Standard Model.  Many searches have been carried out; however, no $5\sigma$ evidence for $CP$ violation has yet been found in these types of decays.  A novel model-independent method for observing $CP$ violation in many-body decays is presented in this paper. It is shown that the sensitivity of this method is significantly larger than those used to-date.  
\end{abstract}


\maketitle

\section{Introduction}

Charge-parity ($CP$) violation is permitted within the Standard Model of particle physics during certain quark-flavor-changing processes.  Such processes are described in the Standard Model by the Cabibbo-Kobayashi-Maskawa (CKM) matrix~\cite{ref:cabibbo,ref:km}.  In many cases, $CP$ violation is suppressed within the CKM picture to the extent that it would not be observable by any current experiment.  New (undiscovered) physical processes and/or particles could provide additional sources of $CP$ violation; thus, simply observing $CP$ violation in many cases would be strong evidence for the existence of physics beyond the Standard Model.

Observing $CP$ violation involves measuring an asymmetry between the decay rate of a process and its $CP$-conjugate (c.c.). 
This asymmetry is the manifestation of the change of sign of a weak phase under charge conjugation.  This effect becomes observable via interference between amplitudes that do and do not contain this weak phase (the amplitudes must also have a non-zero strong phase difference).  

In many-body (three or more daughter particles) decays, $CP$ violation can not only produce an asymmetry in the integrated yields but also in the kinematic distributions of the daughter particles ({\em e.g.}, an asymmetry in the Dalitz plots of a decay and its c.c.).  The presence of resonances in many-body decays assures the existence of non-zero strong phases and many extensions to the Standard Model provide the required large weak phase (see, {\em e.g.}, Refs.~\cite{ref:cpvth1,ref:cpvth2,ref:cpvth3}).  Both model-independent and model-dependent methods have been used to search for $CP$ violation in many-body decays; however, no $5\sigma$ evidence has been found to-date.  Of particular interest in this paper are those processes for which the Standard Model predicts a level of $CP$ violation that is too small to observe in current experiments ({\em e.g.}, many $D \to hhh$ decays, where $h = K$ or $\pi$, satisfy this criteria).  A model-independent observation of $CP$ violation in such processes would be sufficient to establish the existence of physics beyond the Standard Model (for a more detailed discussion, see Ref.~\cite{ref:miranda}).

Studying $CP$ violation in many-body decays is an important part of the LHC$b$ physics program~\cite{ref:lhcb}.  With LHC$b$ now taking data, it is vital that the best possible tools for observing $CP$ violation in many-body decays are available to enhance the sensitivity to physics beyond the Standard Model.  In this paper, I present a novel method for observing $CP$ violation in many-body decays.  In Section~\ref{sec:toy} I describe the physics model used in this study, while in Section~\ref{sec:meth} I give an overview of the method.  Results obtained using current methods and the novel method presented in this paper are given in sections \ref{sec:chi2} and \ref{sec:energy}, respectively.  A summary is presented in Section~\ref{sec:conc}.

\section{Toy-Model Analysis}
\label{sec:toy}
In this paper I consider the decay $X \rightarrow a b c$, where $m_X = 1$ and $m_a = m_b = m_c = 0.1$ are the particle masses (in some units).  All four particles are pseudo-scalars; {\em i.e.}, they all have a spin-parity of $0^-$.  I have chosen to use a three-body decay because these are the most commonly used many-body decays when searching for $CP$ violation.  The method presented in Section~\ref{sec:meth}, however, is not restricted to usage in three-body decays; it is straightforward to apply it to any many-body decay.

The base Dalitz-plot model, which contains no $CP$ violation, is the model used in Ref.~\cite{ref:gof}.  It is constructed using the isobar formalism as follows:
\begin{equation}
  \label{eq:isobar}
  {\cal M}(\vec{x}) = a_{\rm nr} e^{i\phi_{\rm nr}} + \sum_r a_r e^{i\phi_r} \mathcal{A}_r(\vec{x}). 
\end{equation}
In Eq.~\ref{eq:isobar}, $\vec{x} = (m^2_{ab},m^2_{ac})$ represents the position in the Dalitz plot and $a e^{i\phi}$ describes the complex amplitude for each component.  The non-resonant term is denoted by ${\rm nr}$ and is taken to be constant across the Dalitz plot.  The resonant amplitudes, denoted by $\mathcal{A}_r(\vec{x})$, contain contributions from Blatt-Weisskopf barrier factors~\cite{Blatt}, relativistic Breit-Wigner line shapes to describe the propagators and spin factors obtained using the Zemach formalism~\cite{ref:zemach}.  Evaluation of the amplitudes is done using the {\tt qft++} package~\cite{Williams:2008wu}. The resonance properties and fit fractions are shown in Table~\ref{tab:resonance-params}.  The probability density function (p.d.f.) for such a process is easily obtained from the total amplitude as $f(\vec{x}) = |{\cal M}(\vec{x})|^2/\int |{\cal M}(\vec{x})|^2 d\vec{x}$, where the normalization to unity is explicit.

\begin{table}
  \begin{center}
  \begin{tabular}{ccccc}
    \hline
    Daughters & $J^P$ & Mass & Width & Fit Fraction \\
    \hline
    $a,b$ & $0^+$ & 0.3 & 0.025 & 6\% \\
    $a,b$ & $2^+$ & 0.6 & 0.05 & 2\% \\
    $a,c$ & $1^-$ & 0.4 & 0.04 & 18\% \\
    $a,c$ & $0^+$ & 0.7 & 0.1 & 43\% \\
    $b,c$ & $1^-$ & 0.35 & 0.01 & 10\% \\
    $b,c$ & $0^+$ & 0.75 & 0.02 & 17\% \\
    $a,b,c$ & \multicolumn{3}{c}{non-resonant} & 1\% \\
    \hline
  \end{tabular}
  \caption{
    \label{tab:resonance-params}
    Resonances included in the Dalitz-plot model used in this paper (parameter values are for the $CP$-conserving version of the model).
  }
  \end{center}
\end{table}

A variation of this model that contains a moderate amount of $CP$ violation is also considered in this paper.  The $CP$ violation is limited to the $J^P = 1^-$ resonance in the $ac$ system, which has an 18\% fit fraction, and is taken to be
\begin{equation}
  \frac{\Delta a_{1^-_{ac}}}{a_{1^-_{ac}}} = 1.05, \qquad \Delta \phi_{1^-_{ac}} = 10^{\circ},
\end{equation}  
{\em i.e.}, the magnitude and phase of this resonance in the $CP$-violating model for the decay and its c.c.\ are $(a \pm \Delta a/2)$ and $(\phi \pm \Delta \phi/2)$, respectively.  The sample sizes used in these types of analyses tend to be in the $\mathcal{O}(10^3-10^5)$ range (although, $D$ decay samples with $\mathcal{O}(10^6)$ events will soon be available at LHC$b$).  In this paper I will consider sample sizes of $10^4$ events.  Clearly, smaller (larger) amounts of $CP$ violation would be observable with larger (smaller) sample sizes.

\begin{figure*}
  \centering
  \includegraphics[width=0.32\textwidth]{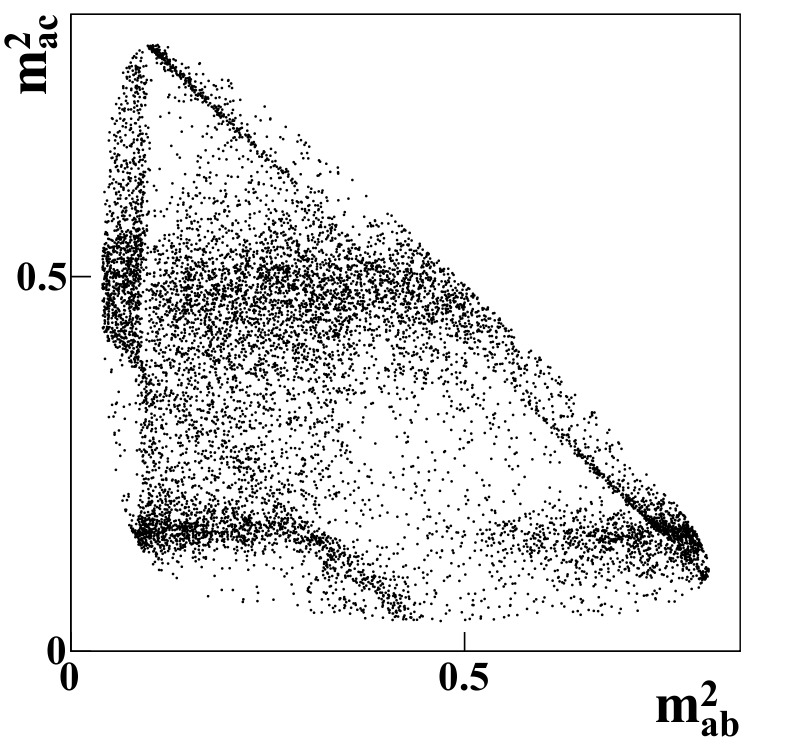} \hspace{0.2in}
  \includegraphics[width=0.32\textwidth]{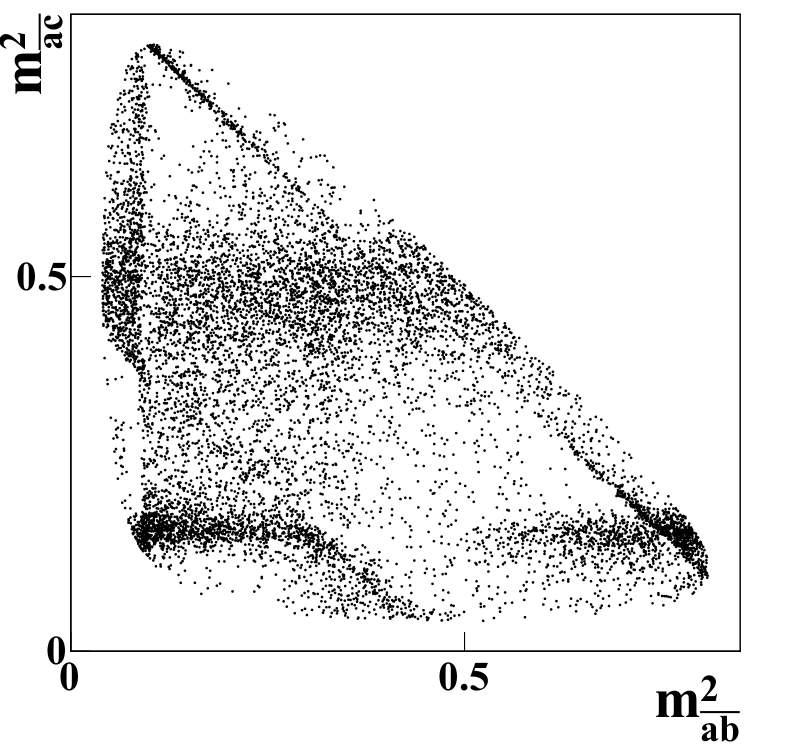}
  \caption[]{\label{fig:dps}
    Example $X \to abc$ (left) and c.c.\ (right) Dalitz plots from the $CP$-violating model. 
  }
\end{figure*}

Ensembles of 100 data sets each for the $CP$-conserving and $CP$-violating versions of the model are produced.  Figure~\ref{fig:dps} shows an example of a single data set produced from the $CP$-violating version of the model.  The integrated direct $CP$ asymmetry, {\em i.e.}, the $CP$ asymmetry in the event yields, is $\sim$ 2\%.  In this paper I will assume that the production asymmetry is not known well enough to use this information; thus, I will only look for $CP$ asymmetry in the decay distributions and not in the event yields.  It is trivial to modify the method presented in Section~\ref{sec:meth} to incorporate the event yields if the production asymmetry is known to a high level of precision.

\section{Method}
\label{sec:meth}

In this section I will describe the method presented in this paper for observing $CP$ violation in many-body decays; however, prior to this, I will first review the methods that have previously been used to search for $CP$ violation in these decays.  
In the absence of $CP$ violation, the decay $X \to abc$ and its c.c.\ have the same parent distribution; thus, a two-sample comparison test between the $X \to abc$ and c.c.\ data sets can be used to observe $CP$ violation in a model-independent way.  Many analyses have used a two-sample binned $\chi^2$ test for this purpose (see, {\em e.g.}, Ref.~\cite{ref:babar1}) but none have observed the {\em golden} $5\sigma$ significance.  Recently another binned approach has been proposed that also provides a useful visualization tool~\cite{ref:miranda}.

Analyses have been performed that have avoided binning the data by employing an unbinned likelihood fit of an isobar model to the data (see, {\em e.g.}, Ref.~\cite{ref:babar1}).  Any significant difference in the resonance amplitude parameters ($a_r$ and $\phi_r$ in Eq.~\ref{eq:isobar}) obtained from the fits to the decay and its c.c.\ could be evidence of $CP$ violation.  The advantage here is that, if the model is accurate, the $CP$-violating parameters can be extracted from the data; however, if the goal of the analysis is to first search for evidence of $CP$ violation and to quote a significance for the observation, then this approach is not optimal due to its introduction of model-dependence into the systematic uncertainties.  There are also some subtleties that need to be accounted for when attempting to quote a significance that are discussed in Section~\ref{sec:res}.

The novel idea presented in this paper is to instead perform an unbinned two-sample test on the data obtained for $X \to abc$ and c.c.\ decays.   In Section~\ref{sec:energy-results} it will be shown that this approach has the following benefits: increased sensitivity to $CP$-violating effects relative to binned methods and no model dependence or any other artifacts that make determining the statistical significance of a $CP$-violation observation difficult.  It is somewhat surprising that this is, in fact, a novel idea and I hope that by presenting it I can also inspire the usage of such techniques in other high energy physics analyses.

The following test statistic correlates the difference between the $X \to abc$ and c.c.\ p.d.f.s, denoted by $f(\vec{x})$ and $\bar{f}(\vec{x})$, respectively, at different points in the multivariate space~\cite{ref:baringhaus,ref:aslan}:
\begin{eqnarray}
  \label{eq:energy-def}
  T &=& \frac{1}{2}\int\int \left(f(\vec{x}) - \bar{f}(\vec{x}) \right)  \left(f(\vec{x}^{\prime}) - \bar{f}(\vec{x}^{\prime}) \right)  \nonumber \\
  && \hspace{1.5in} \times \psi(|\vec{x} - \vec{x}^{\prime}|) d\vec{x} d\vec{x}^{\prime} \nonumber \\
  &=& \frac{1}{2}\int\int \left[f(\vec{x})f(\vec{x}^{\prime}) + \bar{f}(\vec{x})\bar{f}(\vec{x}^{\prime}) -2f(\vec{x})\bar{f}(\vec{x}^{\prime})\right] \nonumber \\ 
  && \hspace{1.5in} \times \psi(|\vec{x}-\vec{x}^{\prime}|) d\vec{x}d\vec{x}^{\prime},
\end{eqnarray}
where $\psi(|\vec{x} - \vec{x}^{\prime}|)$ is a weighting function.  $T$ can be estimated without the need for any knowledge about the forms of $f$ and $\bar{f}$ using $X \to abc$ and c.c.\ data as
\begin{eqnarray}
  \label{eq:t-calc}
  T \approx \frac{1}{n(n-1)}\sum\limits_{i,j>i}^{n} \psi(\Delta\vec{x}_{ij}) \hspace{1.5in}\nonumber \\
  \hspace{0.3in}  + \frac{1}{\bar{n}(\bar{n}-1)}\sum\limits_{i,j>i}^{\bar{n}} \psi(\Delta\vec{x}_{ij}) 
  - \frac{1}{n \bar{n}}\sum\limits_{i,j}^{n,\bar{n}} \psi(\Delta\vec{x}_{ij}),
\end{eqnarray}
where $\Delta\vec{x}_{ij} =|\vec{x}_i - \vec{x}_j|$ and $n$ ($\bar{n}$) is the number of $X \to abc$ (c.c.) events. {\em N.b.}, in the order in which they appear in Eq.~\ref{eq:t-calc}, the sums are over pairs of $X \to abc$ events, pairs of c.c.\ events and pairs consisting of an $X \to abc$ event and a c.c.\ event, respectively.  Eq.~\ref{eq:t-calc} is simply Eq.~\ref{eq:energy-def} rewritten using the standard Monte Carlo integration approximation, along with the fact that $\int f(\vec{x})d\vec{x} = \int \bar{f}(\vec{x}) d\vec{x}= 1$.

It is straightforward to calculate $T$ using Eq.~\ref{eq:t-calc} once a metric is chosen that defines distance in the multivariate space (see Ref.~\cite{ref:gof} for a detailed discussion on metrics; this choice has almost no effect on the results). It is worth noting here that the larger the difference is between $f$ and $\bar{f}$ the larger the expectation value of $T$ becomes; thus, $T$ can be used to determine the goodness of fit (g.o.f.) of the data to the hypothesis $f = \bar{f}$ ({\em i.e.}, no $CP$ violation).  

This method is referred to as the {\em energy test} in Ref.~\cite{ref:aslan} due to the fact that if $\psi(x) = 1/x$ then Eq.~\ref{eq:energy-def} is the electrostatic energy of two charge distributions of opposite sign. Ref.~\cite{ref:aslan} also notes that the electrostatic energy is minimized if the charges neutralize each other, {\em i.e.}, if $f = \bar{f}$.  The choice of weighting function plays an analogous role to that of the bin width and binning scheme in a $\chi^2$ test. In Ref.~\cite{ref:gof} it was found that for Dalitz-plot analyses a gaussian weighting function is optimal when comparing a data set to a p.d.f.  In that scenario a Monte Carlo data set is sampled from the p.d.f.\ and generated with great enough statistics that fluctuations within the sample are negligible.  This permits the determination of the properties of the weighting function based on the p.d.f.\ (physics) and not on the size of the data sample.  

The problem being studied in this paper is different in that it involves comparing two data sets; thus, one cannot simply increase the statistics of either sample.  Because of this, the sample sizes must factor into the weighting function in some way (at least, for samples of the sizes used in this study).  The following weighting functions both work equally well for this analysis:
\begin{eqnarray}
  \label{eq:psi-def}
  \psi(\Delta\vec{x}_{ij}) &=& -\log{(\Delta\vec{x}_{ij} + \epsilon)}, \\
    \psi(\Delta\vec{x}_{ij}) &=& e^{-\Delta\vec{x}_{ij}^2/2 \sigma^2}, 
\end{eqnarray} 
where $\epsilon$ is of the order of $[|f(\vec{x})|_{\rm max}(n+\bar{n})]^{-1}$ and $\sigma$ is of the order of the mean distance to the $k^{th}$ nearest neighbor (I chose $k =100$) in the sample.  The results obtained with both weighting functions are consistent.  The results below were obtained using Eq.~\ref{eq:psi-def}.  The maximum value of the p.d.f., $f_{\rm max}$, can be estimated using the local density near each event.   The exact value used for $\epsilon$ is not important. I varied it by an order of magnitude in both directions and obtained consistent results (in fact, there was almost no change in the $T$-values obtained).

The distribution of $T$ for the case where $f = \bar{f}$ is not known; thus, to convert the $T$-value into a $p$-value the permutation test~\cite{ref:fisher} (or another resampling method) must be used.  This involves making pseudo data sets by randomly assigning the labels ``decay'' and ``c.c.\ decay'' to each event such that there are $n$ ``decay'' events and $\bar{n}$ ``c.c.\ decay'' events in each pseudo data set.  $T$ is then calculated with these designations for each event.  This process is repeated $n_{\rm perm}$ times to obtain the set of values $\{T_1 \ldots T_{n_{\rm perm}}\}$.  The $p$-value is then the fraction of elements in the set that are larger than the $T$-value obtained using the measured event designations.  For a detailed discussion on this topic, see Ref.~\cite{ref:good}.

\section{Binned Results}
\label{sec:chi2}
The normalized two-sample $\chi^2$ statistic is given as follows:
\begin{equation}
  \label{eq:chi2}
  \chi^2 = \sum\limits_{i=1}^{n_b} \frac{( o_i \bar{n} - \bar{o}_i n)^2}{n\bar{n}(o_i + \bar{o}_i)},
\end{equation} 
where $o_i$ and $\bar{o}_i$ are the observed number of events in the $i^{th}$ bin in each of the data sets and $n_b$ is the number of bins.  The test statistic defined in Eq.~\ref{eq:chi2} will approximately follow a $\chi^2$ distribution with $n_b -1$ degrees of freedom for any {\em reasonable} choice of binning scheme; {\em i.e.}, given that the bins are chosen such that there aren't too many low-occupancy bins.  I have chosen to use 500 bins in the allowed Dalitz space which yields an average of 20 events per bin (the number of low-occupancy bins is less than 10\%).  In Section~\ref{sec:sig} I will discuss in detail the determination of significance using this method.  For now, I will simply proceed under the assumption that my reasonable choice of bins allows me to determine $p$-values using {\tt TMath::Prob}~\cite{ref:tmath} (a standard assumption in high energy physics).

\begin{figure}
  \centering
  \includegraphics[width=0.32\textwidth]{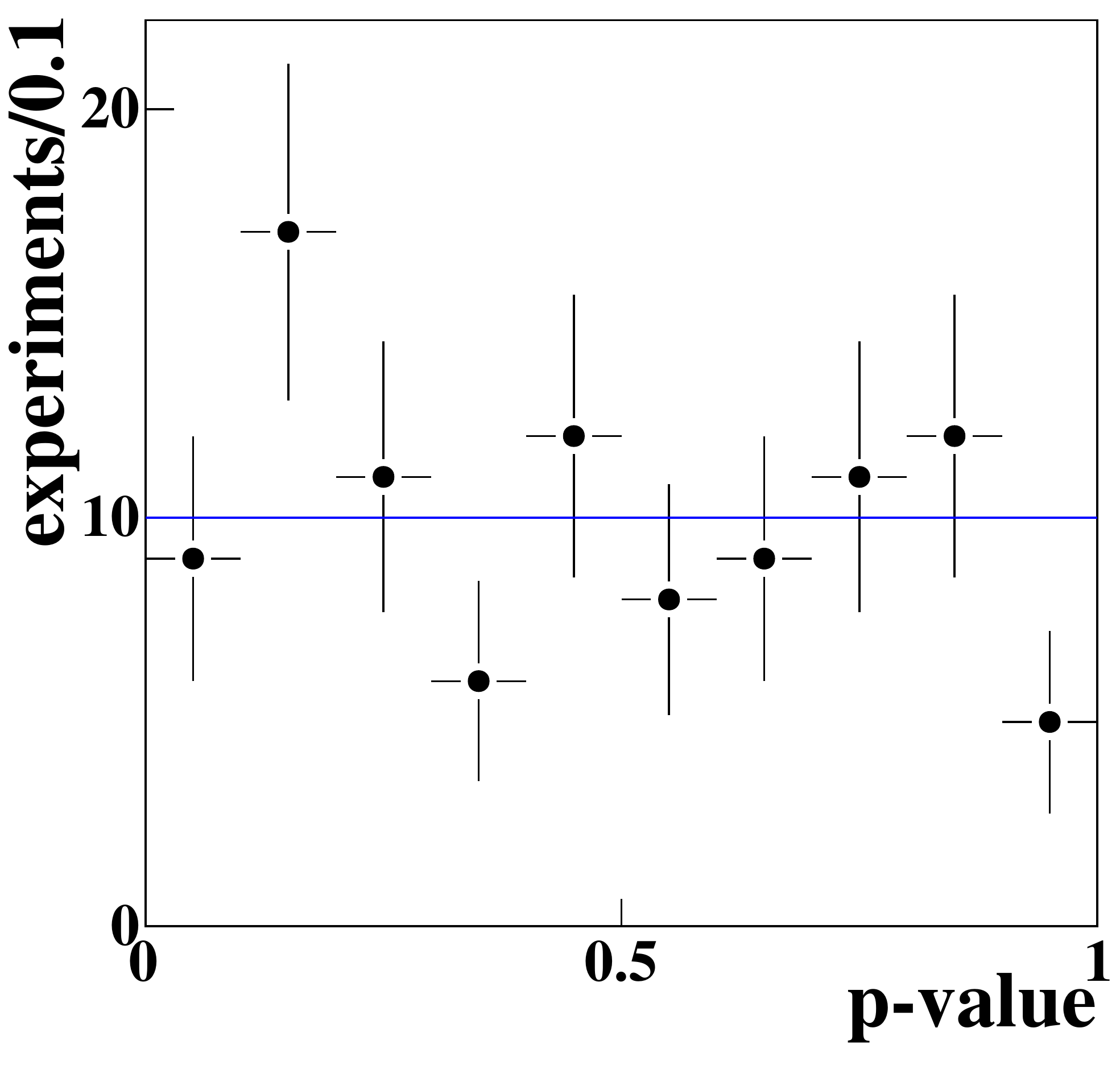}
  \caption[]{\label{fig:p-chi2}
    $p$-value distribution obtained using the $\chi^2$ method on the $CP$-violating ensemble of data sets.  The blue line shows the expected distribution for the $CP$-conserving case.
  }
\end{figure}

Figure~\ref{fig:p-chi2} shows the $p$-value distribution obtained by testing the $CP$-conserving hypothesis using the $CP$-violating ensemble of data sets.  The results are consistent with the hypothesis (even though the data does violate $CP$ symmetry).  Table~\ref{tab:sig-dev} shows the fraction of data sets that exceed the one, two and three $\sigma$ levels of significance.  These results are also what would be expected if the two data sets did, in fact, share a parent distribution (no $CP$ violation).  From these results I conclude that the $\chi^2$ test is unable to detect $CP$ violation at the level I have introduced it into my model Dalitz-plot analysis.

\begin{figure}
  \centering
  \includegraphics[width=0.32\textwidth]{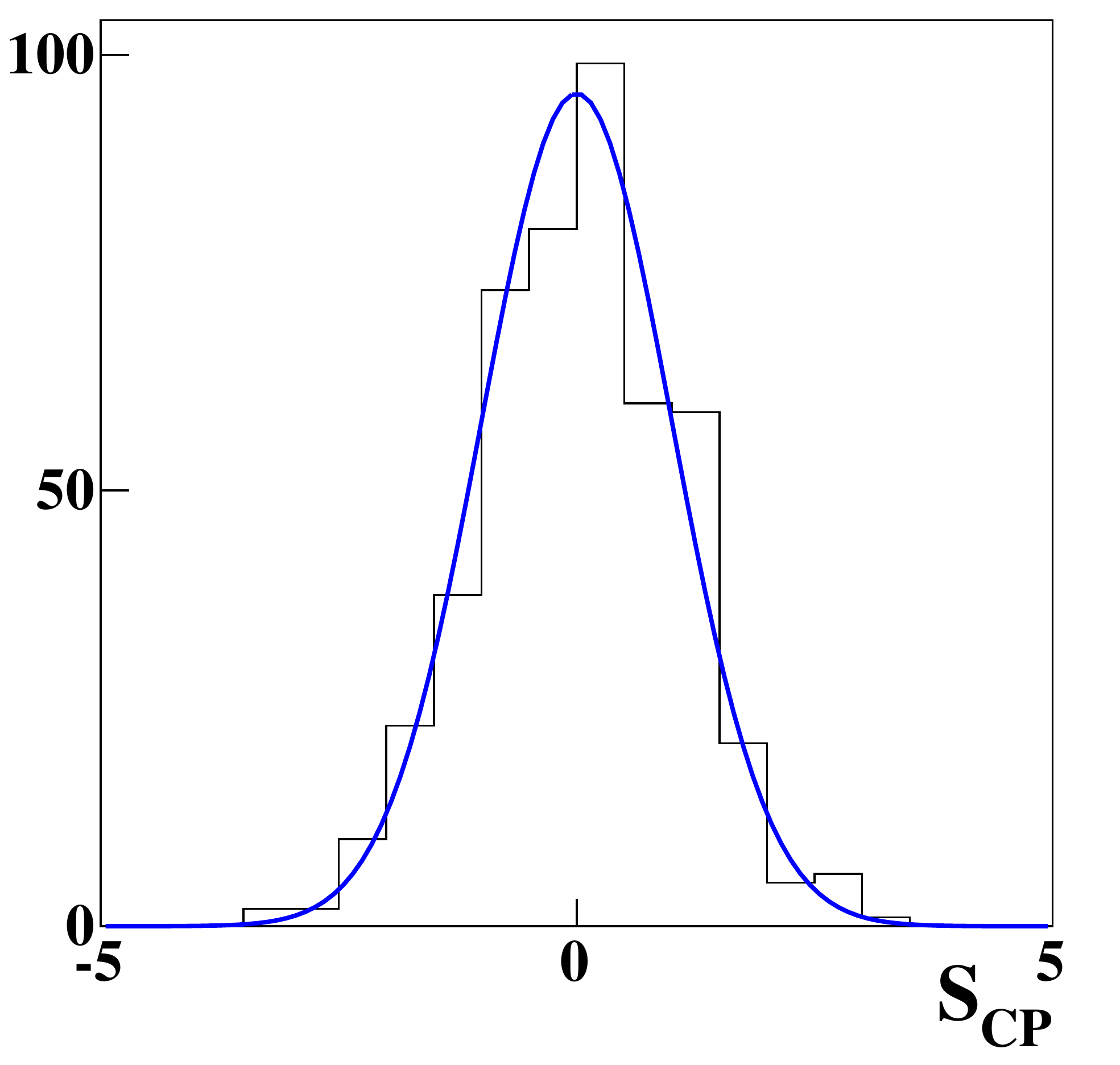}
  \caption[]{\label{fig:miranda}
    The {\em mirandized} distribution from a $CP$-violating data set.  The blue line shows a standard normal gaussian.
  }
\end{figure}

Ref.~\cite{ref:miranda} argues that rather than calculating the test statistic defined in Eq.~\ref{eq:chi2}, one should instead plot the distribution obtained by calculating the following quantity for each Dalitz-plot bin:
\begin{equation}
  S_{CP}^i = \frac{o_i - \bar{o}_i}{\sqrt{o_i + \bar{o}_i}}.
\end{equation}
For the analysis in this paper an additional factor needs to be applied to normalize the two data sets since I am assuming that the production asymmetry is not known.  Figure~\ref{fig:miranda} shows the so-called {\em mirandized} distribution for a $CP$-violating data set that exhibits a $2\sigma$ deviation in the $\chi^2$ test.  The expected (standard normal gaussian) distribution under the $CP$-conserving hypothesis is also shown; there is no obvious discrepancy.  The $p$-values obtained using this method are found to be consistent with those obtained using the $\chi^2$ test; thus, I conclude that this method is also not sensitive enough to detect $CP$ violation at the level contained in my model Dalitz-plot analysis.

\section{Unbinned Multivariate Results}
\label{sec:energy}
\subsection{Energy Test Results}
\label{sec:energy-results}
\begin{figure}
  \centering
  \includegraphics[width=0.32\textwidth]{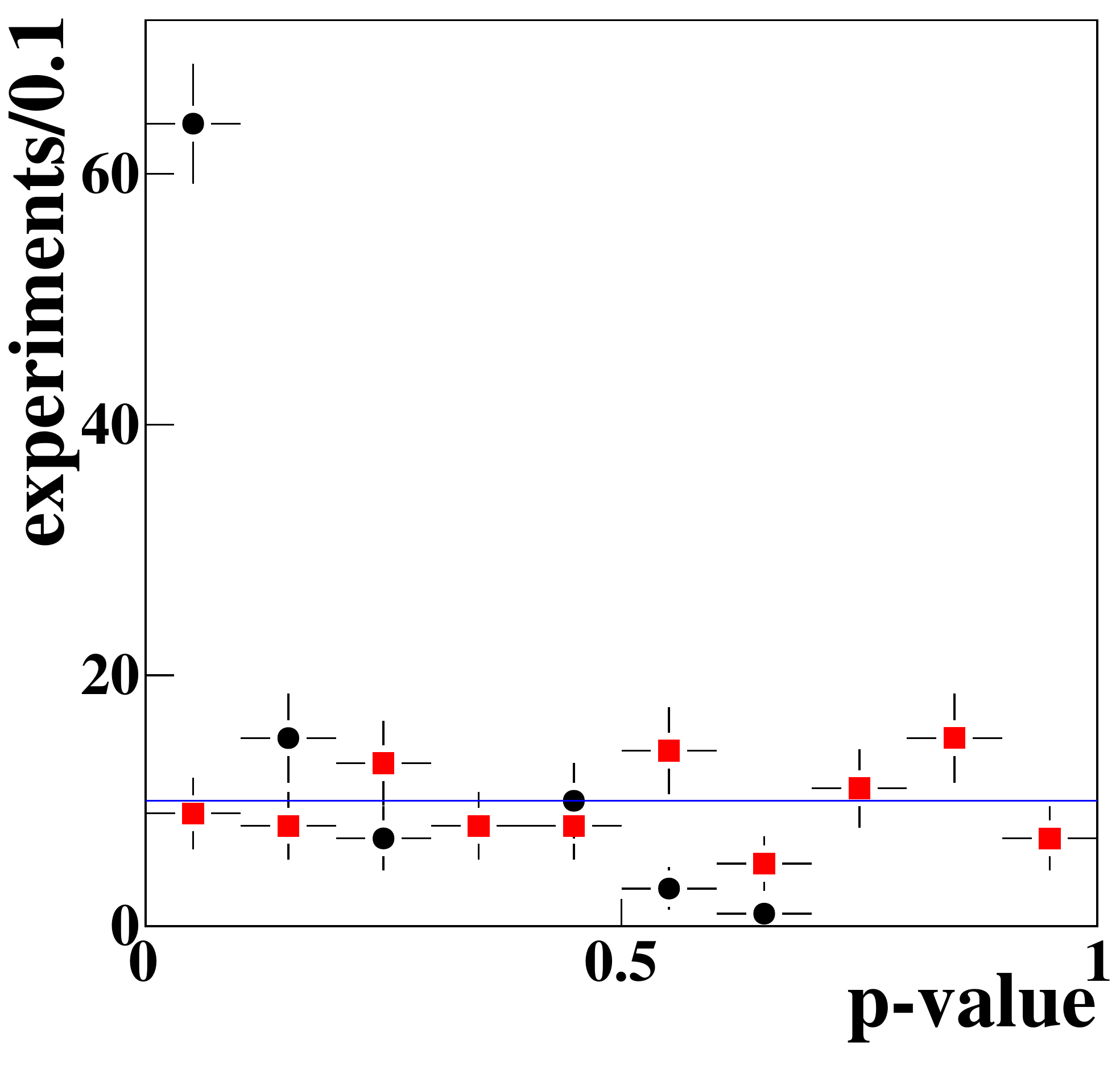}
  \caption[]{\label{fig:t-p}
    $p$-value distributions obtained using the energy test on the $CP$-conserving (red squares) and $CP$-violating (black circles) ensembles of data sets.  The (solid blue) line shows the expected distribution for the $CP$-conserving case.
  }
\end{figure}

I will first demonstrate that the energy test produces the expected results for the $CP$-conserving case.  Figure~\ref{fig:t-p} shows the $p$-value distribution obtained by calculating $T$ for each $CP$-conserving data set using Eq.~\ref{eq:t-calc} and then converting this into a $p$-value using the permutation test.   The $p$-value distribution is consistent with being uniform as expected.  

The same procedure is then applied to the $CP$-violating ensemble of data sets; the resulting $p$-value distribution is also shown in Fig.~\ref{fig:t-p}.  There is clear evidence of disagreement between the data and the $CP$-conservation hypothesis.  In fact, two-thirds of all data sets permit rejection of the $CP$-conserving hypothesis at the 90\% confidence level.  Table~\ref{tab:sig-dev} shows the fraction of data sets that exceed the one, two and three $\sigma$ levels of significance.  These results are very impressive when compared to those obtained using the $\chi^2$ test.  The energy test yields approximately 1:1 and 1:7 odds of observing two and three $\sigma$ significances, respectively, for the $CP$-violating Dalitz-plot model under study in this paper.  

These odds will increase if either the level of $CP$-violation in the model or the size of the samples is increased (or both).  The performance of the $\chi^2$ test will also improve in these circumstances.  Given the broad range of many-body decays that can be used to look for evidence of $CP$-violation, it is not possible to fully map out the performance statistics for all possible situations. The important result here is that the energy test vastly out-performs the $\chi^2$ test under what are typical conditions encountered in these types of analyses (more discussion on this topic can be found in Section~\ref{sec:conc}). 

\begin{table}
  \begin{center}
  \begin{tabular}{c|ccc}
    \hline
    test & $1\sigma$(\%) & $2\sigma$(\%) & $3\sigma$(\%) \\
    \hline
    $\chi^2$ &  38$\pm$5 & 3$\pm$2 & 0$\pm$1   \\
    energy & 87$\pm$3 &  52$\pm$5 & 13$\pm$3  \\
    \hline
  \end{tabular}
  \caption{
    \label{tab:sig-dev}
    Observed deviation levels for the $CP$-violating ensemble of data sets for the $\chi^2$ and energy tests.
  }
  \end{center}
\end{table}

\subsection{Visualization}

Once the significance level of $CP$ violation has been determined, the analyst may want to determine which regions of the Dalitz plot exhibit discrepancies with the $CP$-conserving hypothesis.  Refs.~\cite{ref:baringhaus,ref:aslan} do not provide any such tools; however, it is not difficult to invent one.  Consider the contribution from each $X \to abc$ event to $T$ which I will define as follows:
\begin{equation}
  \label{eq:ti-def}
  T_i = \frac{1}{2n(n-1)}\sum\limits_{j \neq i}^n \psi(\Delta \vec{x}_{ij}) - \frac{1}{2n\bar{n}}\sum\limits_{j}^{\bar{n}} \psi(\Delta \vec{x}_{ij}), 
\end{equation}
where the first sum is over $X \to abc$ events and the second sum is over c.c.\ events.  {\em N.b.}, an equation for events from the c.c.\ data set is similarly defined. It is important to realize that the $T_i$'s are not independent due to the {\em interaction potential} $\psi$.  

It is straightforward to assign a significance to the maximum $T_i$ value obtained in a data set, $T_i^{\rm max}$, using the permutation test.  This is done by simply determining the $T_i^{\rm max}$ distribution in the same way as that of $T$ above.  Due to the non-independence of the $T_i$ values this method only permits assigning significance to $T_i^{\rm max}$; however, this is enough to design a simple visualization tool.  

\begin{figure}
  \centering
  \includegraphics[width=0.32\textwidth]{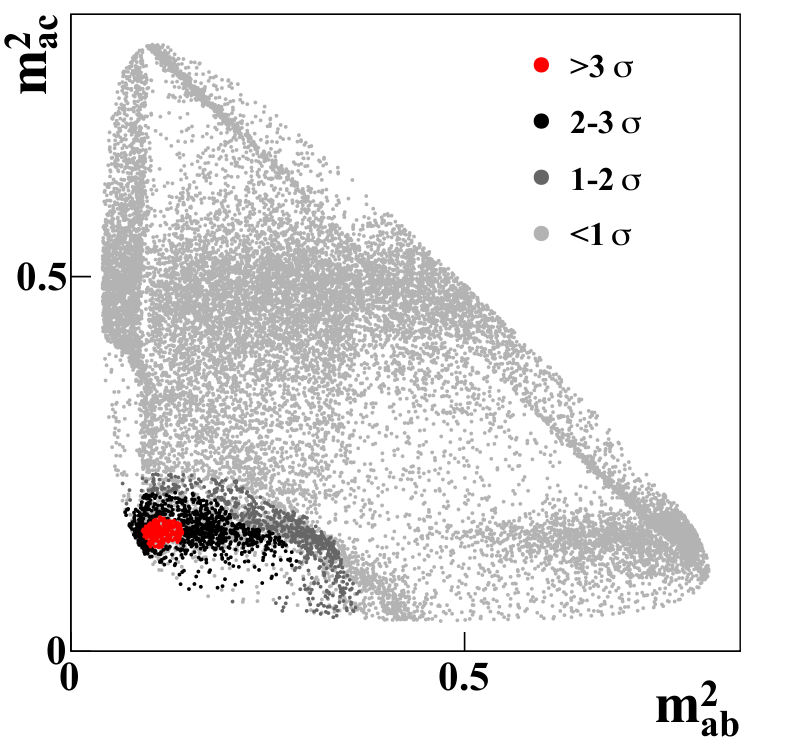}
  \caption[]{\label{fig:t-dp}
    (Color Online) $T$-value distribution in terms of $T_{\rm max}$ significance levels (see text for details).  
  }
\end{figure}

Figure~\ref{fig:t-dp} shows the events in a $CP$-violating data set (from both decays) color-coded according to which $T_i^{\rm max}$ significance band contains their $T_i$ values.  This particular data set exhibited a $3\sigma$ deviation using the energy test and this visualization method allows us to see which regions of the Dalitz plot contribute the most to that significance ({\em n.b.}, this is the same data set used to make Fig.~\ref{fig:miranda}).  I stress again that the $T_i$ values are correlated; thus, the total number of events in each band is not easily interpretable.  What can be said is that the $T_i$ values obtained in most of the Dalitz plot are consistent with the $CP$-conserving hypothesis, but those in the region of overlap between the $J^P = 1^-$ resonance in the $ac$ system and the $J^P = 0^+$ resonance in the $ab$ system are not.  Only 0.3\% of all $CP$-conserving data sets will have at least one event with a $T_i$ value in the $3\sigma$ band (none of the 100 $CP$-conserving data sets used in this paper have any).  While the rigorous $p$-value determination should be left to the energy test itself, this visualization method provides a useful tool for displaying the results.

\subsection{Testing Resonance Regions}
\label{sec:res}
\begin{figure}
  \centering
  \includegraphics[width=0.32\textwidth]{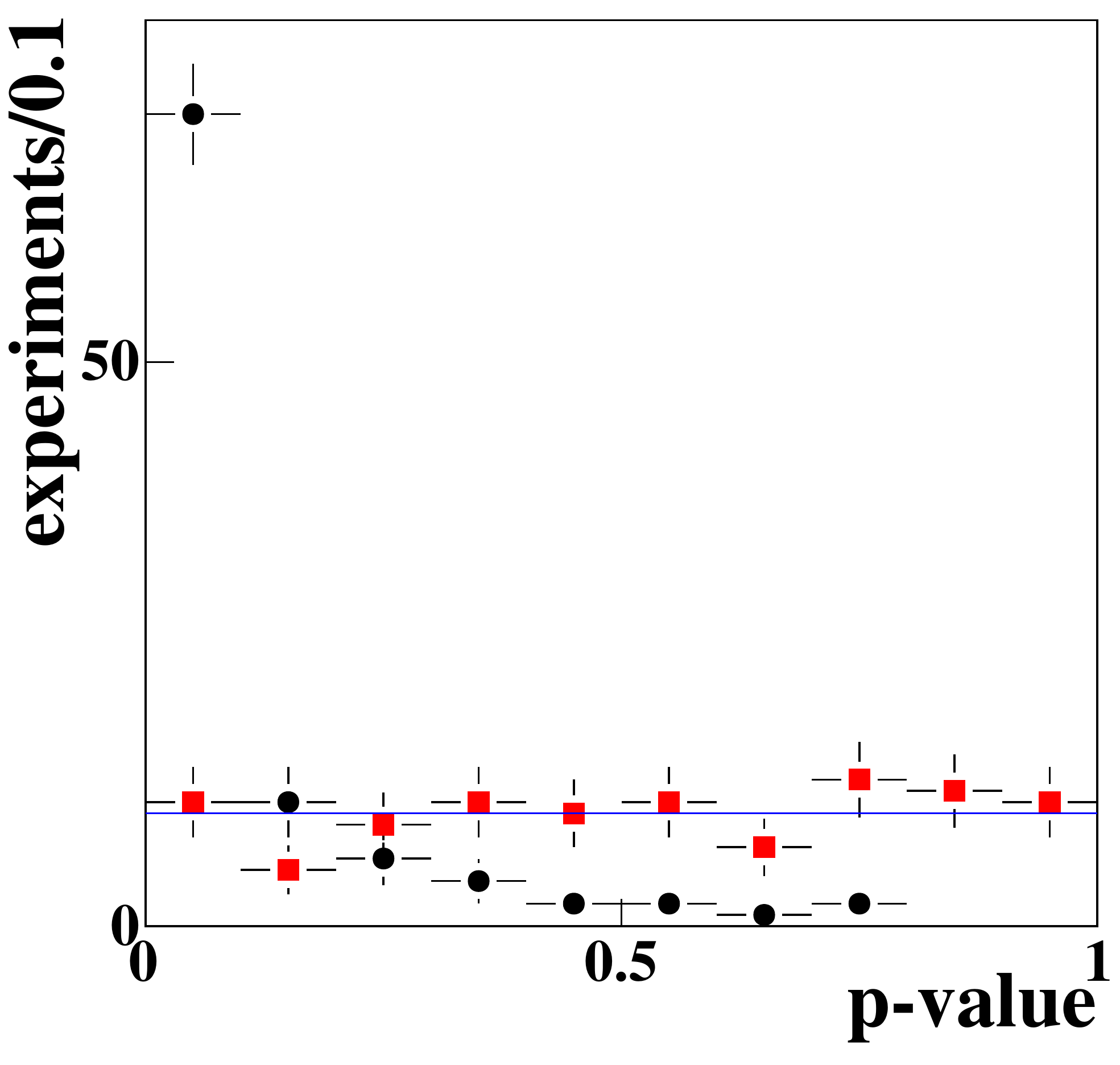}
  \caption[]{\label{fig:ac1-bc0}
    $p$-value distributions obtained from the regions around the $J^P = 1^-$ resonance in the $ac$ system (black circles) and the $J^P = 0^+$ resonance in the $bc$ system for the $CP$-violating ensemble of data sets.
  }
\end{figure}

One could, of course, also test regions of the Dalitz plot instead of the entire Dalitz space.  {\em E.g.}, one could define regions around various resonances and determine whether the data obtained in these regions is consistent with the $CP$-conserving hypothesis.  Figure~\ref{fig:ac1-bc0} shows the $p$-value distributions obtained from the regions around the $J^P = 1^-$ resonance in the $ac$ system and the $J^P = 0^+$ resonance in the $bc$ system (with the small overlap with the previous resonance region removed) for the $CP$-violating ensemble of data sets.  There is no evidence for $CP$ violation in the $bc$ resonance region used here; however, there is a sizable discrepancy in the $ac$ one (this is the resonance that does exhibit $CP$ violation in its parameters in my model).  

Examining individual resonance regions separately could be a viable approach; however, there are two things to keep in mind: (1) if the regions overlap then they are not independent samples; (2) the probability of obtaining a $p$-value less than $\alpha$ in $n_{R}$ resonance regions is not $\alpha$.  These facts mean that some care must be taken when determining the significance of a $CP$-violation observation this way.  I note here that these same issues are also present when using the model-dependent approach described in Section~\ref{sec:meth}.

\subsection{Background \& Efficiency}

Up to this point I have ignored the presence of background events and detector inefficiencies.  In fact, if the relative size and shape of the background is expected to be the same for both the $X \to abc$ and c.c.\ decays then it can be ignored since in this scenario the two distributions would still be expected to be consistent under the $CP$-conserving hypothesis.  This is also true for the detector inefficiency: if the detector efficiency is the same for both decays, then it can be ignored (following the same line of reasoning).  This is true for the $\chi^2$ test as well.

To a good approximation the conditions required to ignore both of these effects will likely be present in many of these types of analysis; however, I will now describe how to alter the energy test to deal with the case where these conditions are not met.  This is simply done by rewriting  Eq.~\ref{eq:t-calc} as follows:
\begin{eqnarray}
  \label{eq:t-calc-bkgd}
  T \approx \frac{1}{W^2}\sum\limits_{i,j>i}^{n}w_iw_j \psi(\Delta\vec{x}_{ij}) \hspace{1.5in}\nonumber \\
+ \frac{1}{\bar{W}^2}\sum\limits_{i,j>i}^{\bar{n}}w_iw_j \psi(\Delta\vec{x}_{ij}) 
 - \frac{1}{W \bar{W}}\sum\limits_{i,j}^{n,\bar{n}} w_iw_j\psi(\Delta\vec{x}_{ij}),
\end{eqnarray}
where $w_i=P^i_S/P^i_D$ are each event's weight factor and account for the signal and detection probabilities, $P_S$ and $P_D$, respectively, and $W$ ($\bar{W}$) is the sum of weight factors for $X \to abc$ (c.c.) events.  The rest of the procedure is unaffected by either of these issues.



\subsection{Large Significance Approximations}
\label{sec:sig}

\begin{figure}
  \centering
  \includegraphics[width=0.32\textwidth]{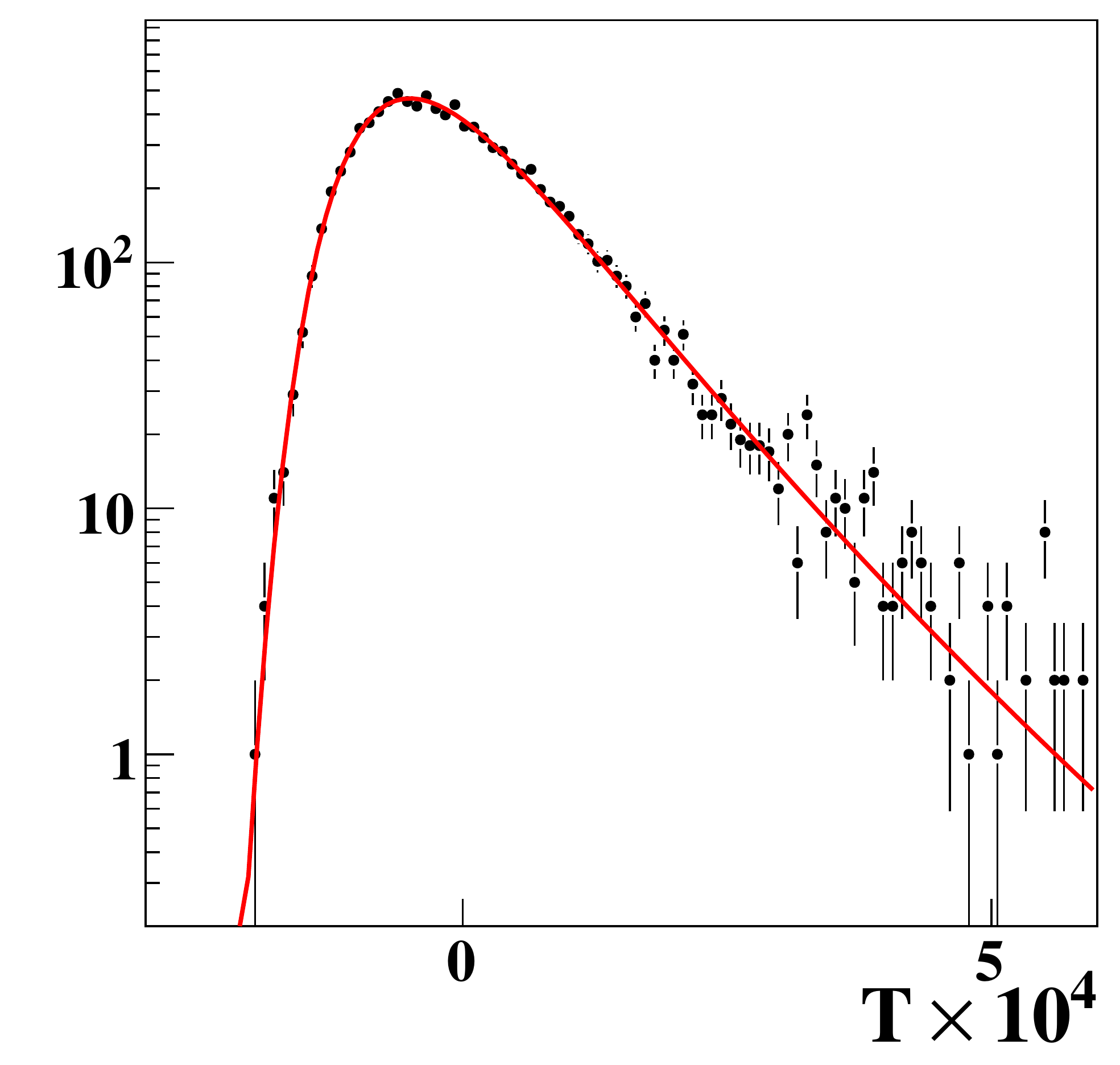}
  \caption[]{\label{fig:t-perm}
    $T$ distribution obtained using the permutation test.  The solid line represents a fit to a generalized extreme value function.
  }
\end{figure}

The permutation test is used to determine the $p$-value for each $T$-value obtained using the energy test.  The downside to this approach is that it requires $n_{\rm perm}$ permutations to determine if $p < 1/n_{\rm perm}$; this can take a lot of CPU time when $n_{\rm perm}$ is large.  Ref.~\cite{ref:aslan} notes that the $T$-value distributions appear to follow the form of a generalized extreme value function (which has three free parameters).  Figure~\ref{fig:t-perm} shows the $T$-value distribution obtained from ten thousand permutations.  Even out to this large a value of $n_{\rm perm}$, the $T$ distribution is still described excellently by the generalized extreme value function; thus, one could generate a smaller number ({\em e.g.}, $n_{\rm perm} = 100$) of permutations and then fit the extreme value function to these $T$-values and estimate the $p$-value using this function (rather than generating more permutations).

The reader may think that this is unacceptable since it introduces some unknown level of uncertainty into the significance calculation; however, there is also an unknown level of uncertainty in the $p$-value determination using the $\chi^2$ test.  The reason for this is that there are a number of requirements that a data set must meet in order for the $\chi^2$ statistic to follow the limiting distribution used to determine the $p$-value.  In practice all of these requirements are never met ({\em e.g.}, $n \neq \infty$) and so the $p$-value obtained is an approximation.  It is typically assumed that this approximation is {\em good enough}; {\em i.e.}, its validity is rarely tested.

The permutation test is referred to as an {\em exact} test because the only uncertainty on the $p$-values it yields is statistical (dependent on $n_{\rm perm}$) and calculable (they are binomial; see, {\em e.g.}, Appendix~C of Ref.~\cite{ref:gof}).  The permutation test can also be used to determine the $p$-values for the $\chi^2$ test.  I have done this and found that for the {\em reasonable} binning scheme chosen in Section~\ref{sec:chi2} the mean discrepancy in the asymptotic $p$-value calculation is $\mathcal{O}(0.1)$.  {\em I.e.}, the $p$-value obtained from {\tt TMath::Prob} is, on average, about 0.1 units larger than the one obtained using the permutation test (although the differences near 0 and 1 are much smaller).

What went wrong?  The answer is that nothing went {\em wrong} but that it is important one realizes that $p$-values obtained in this way are only approximations.  To determine how good the approximation is, one needs to perform the permutation test.  Thus, approximating the $p$-values for the energy test by fitting a smaller number of permutations to an extreme value distribution is no less valid than assuming a test statistic follows a limiting $\chi^2$ distribution.

\section{Summary}
\label{sec:conc}
I have shown that under conditions that are typical when searching for $CP$ violation in many-body decays that an unbinned multivariate two-sample test vastly out performs the binned $\chi^2$ test.  I have also presented a novel method for visualizing the regions of the Dalitz plot that make the largest contributions to the significance.  
I wish to conclude by noting that there is no uniformly most powerful goodness-of-fit test and so there may be situations where other tests perform better than the one explored in this paper or, perhaps, a new test will be invented in the future that out performs all currently available tests.   Physicists should always seek to use, and more importantly understand, the best available tools.

\section*{Acknowledgements}
I am grateful to my colleagues from the LHCb experiment, and would particularly like to thank Ulrik Egede, Tim Gershon and Vladimir Gligorov for discussions.
This work is supported by the Science and Technology Facilities Council (United Kingdom).


\begin{thebibliography}{99}

\bibitem{ref:cabibbo}
  N.~Cabibbo,
  Phys.\ Rev.\ Lett.\  {\bf 10}, 531 (1963).

\bibitem{ref:km}
  M.~Kobayashi and T.~Maskawa,
  Prog.\ Theor.\ Phys.\  {\bf 49}, 652 (1973).


\bibitem{ref:cpvth1} S. Bianco, F.L. Fabbri, D. Benson, and I. Bigi, Riv.\ Nuovo Cim.\ {\bf 26N7}, 1 (2003).

\bibitem{ref:cpvth2} A.A. Petrov, Phys.\ Rev.\ {\bf D69}, 111901 (2004).

\bibitem{ref:cpvth3} Y. Grossman, A.L. Kagan, and Y. Nir, Phys.\ Rev.\ {\bf D75}, 036008 (2007).

\bibitem{ref:miranda} I. Bediaga {\em et al.}, Phys.\ Rev.\ {\bf D80}, 096006 (2009).

\bibitem{ref:lhcb} B. Adeva {\em et al.} [LHC$b$ Collaboration], [arXiv:0912.4179].

\bibitem{ref:gof}
  M. Williams, JINST {\bf 5}, P09004 (2010).

\bibitem{Blatt}
  J.~Blatt and V.~E.~Weisskopf, {\it Theoretical Nuclear Physics}, J.~Wiley, New York (1952).

\bibitem{ref:zemach}  C.~Zemach, 
  Phys.\ Rev.\  {\bf 133} B1201 (1964);  C.~Zemach, 
  Phys.\ Rev.\  {\bf 140} B97 (1965).

\bibitem{Williams:2008wu}
  M.~Williams,
  Comp. Phys. Comm. {\bf 180}, 1847 (2009).

\bibitem{ref:babar1} B. Aubert {\em et al.} [BABAR Collaboration], Phys.\ Rev.\ {\bf D78}, 051102(R) (2008).




\bibitem{ref:baringhaus} L. Baringhaus and C. Franz, 
  J. Multivariate Anal. {\bf 88}, 190-206 (2004).

\bibitem{ref:aslan} B. Aslan and G. Zech, 
  Stat. Comp. Simul. {\bf 75}, Issue 2 109-119 (2004); 
  B. Aslan and G. Zech, 
  Nucl. Instrum. Methods {\bf A537}, 626-636 (2005).

\bibitem{ref:fisher} R.A. Fisher, 
  {\em The Design of Experiments}, 
  Oliver and Boyd Ltd., London (1935).

\bibitem{ref:good} P. Good, 
  {\em Permutation Tests: a Practical Guide to Resampling Methods for Testing Hypotheses}, 
  Springer-Verlag, New York (1994).

\bibitem{ref:tmath} A typical example of an algorithm that calculates limiting $p$-values for the $\chi^2$ test can be found in the ROOT {\tt TMath::Prob} method [root.cern.ch].


\end{thebibliography}
\end{document}